\documentclass[12pt]{article}
\usepackage{graphicx}
\usepackage{amsmath}
\usepackage{amsfonts}
\usepackage{amssymb}

\begin{document}
\begin{titlepage}
\begin{flushright}
CAMS/03-01\\
\end{flushright}
\vspace{.3cm}
\begin{center}
\renewcommand{\thefootnote}{\fnsymbol{footnote}}
{\Large\bf Spontaneous Symmetry Breaking for Massive Spin-2
Interacting with Gravity} \vskip20mm {\large\bf{Ali  H.
Chamseddine \footnote{email: chams@aub.edu.lb}}}
\renewcommand{\thefootnote}{\arabic{footnote}}
\vskip2cm {\it Center for Advanced Mathematical Sciences (CAMS)
and\\
Physics Department, American University of Beirut, Lebanon.\\}
\end{center}
\vfill
\begin{center}
{\bf Abstract}
\end{center}
\vskip.3cm An action for a massless graviton interacting with a
massive tensor field is proposed. The model is based on coupling
the metric tensor to an $SP(4)$ gauge theory spontaneously broken
to $SL(2,C)$. The symmetry breaking is achieved by using a Higgs
multiplet containing a scalar field and a vector field related by
a constraint. We show that in the non-unitary gauge and for the
Fierz-Pauli form of the mass term, the six degrees of freedom of
the massive tensor are identified with  two tensor helicities, two
vector helicities of the Goldstone vector, and two scalars present
in the Goldstone multiplet. The propagators of this system are
well behaved, in contrast to the system consisting of two tensors.
\end{titlepage}

\section{\bigskip Introduction}

It is well known that the action for massive spin-2 field does not have a
smooth limit to the massless case \cite{vv}, \cite{za} and there are many
doubts in the literature about the consistency of this theory \cite{bd},
\cite{dps}, \cite{nw}. The Fierz-Pauli mass term is required to give mass to
the five dynamical degrees of freedom expected for a massive spin-2 field
\cite{fp} . In a covariant formulation, the massive spin-2 field is
represented, like the graviton, by a symmetric tensor. However, gauge freedom
associated with diffeomorphism invariance guarantees that the graviton is
massless. This means that there are six degrees of freedom associated with the
symmetric tensor for the massive spin-2 field, as the four gauge degrees of
freedom resulting from diffeomorphisms, having been used for the graviton, are
not available. The Fierz-Pauli choice for the mass term decouples the
independent sixth scalar degree of freedom from the five massive spin-2
degrees of freedom, only at the linearized level \cite{bd}, \cite{dk}. Quantum
corrections do not preserve the Fierz-Pauli form of the mass term, and implies
that the sixth degree of freedom will have ghost interactions \cite{ss}. The
renewed interest in massive spin-2 fields (also referred to as massive
gravitons) comes from different directions, mainly from brane models where a
different metric is taken on each of the branes \cite{kmprs}, \cite{grs},
\cite{kr}. As there is only one diffeomorphism to be preserved by the full
system, only one combination of the metric tensors can be associated with the
massless graviton. The other combination(s) will not have diffeomorphism
invariance and will (each) correspond to six degrees of freedom. \ The mixing
of the metrics is governed by the Fierz-Pauli choice, which guarantees that
the sixth scalar mode decouples at the linearized level. \ When quantum
corrections are taken into account, one finds that the scalar mode acquires a
kinetic term with ghosts. \ The propagator of the massive field does not have
a smooth $\ m\rightarrow0$ limit, and suffers from the Dam-Veltman
discontinuity \cite{vv}, \cite{za}. Recently, it was shown by Arkani-Hamed,
Georgi and Schwartz \cite{ags}, that the pathological behavior of the massive
graviton propagator could be regulated by spontaneously breaking the symmetry
of the two diffeomorphisms. This was done by introducing a Goldstone vector
field which maps the coordinates of one brane to the other. This is similar to
the use of Stueckelberg field for the description of massive gravity
\cite{siegel}.

The idea in \cite{ags} is closely related to an approach proposed many years
ago by Chamseddine, Salam and Strathdee \cite{css} for strong gravity
\cite{iss}. The system considered in \cite{css} is based on the gauge symmetry
$SP(4)\times SP(4)$ broken spontaneously to $SL(2,C)$ through a Higgs field
transforming under both gauge groups\footnote[1]{In \cite{css} a supergravity
system based on gauging the graded algebra $OSP(1;4)\times OSP(1;4)$ was
considered. Here we shall only consider the bosonic version.}. It was shown
that in the unitary gauge a Fierz-Pauli term does arise for one combination of
the gauge fields associated with the massive graviton, while the other
combination gave a massless graviton. The interactions of the Higgs field
could be evaluated by working in the non-unitary gauge where the Goldstone
mode is kept. The work in \cite{css} depends on the idea that the
gravitational field could be formulated as a gauge theory of the $SP(4)$ gauge
group where the spin-connection is taken as the gauge field of the $SL(2,C)$
subgroup of $SP(4).$ The vierbein is the gauge field of the four remaining
generators of $SP(4)$ \cite{cho}, \cite{ch}. The massive graviton was tuned to
have a very heavy mass and very strong coupling. This arrangement is not
necessary, and it is now preferable to have a very light graviton with weak
coupling. Although this formulation is very elegant, the Higgs structure in
the non-unitary gauge is complicated making the analysis of the Goldstone
multiplet non transparent. The complication is due to the fact that no
space-time metric is used but instead gauge fields are introduced, which then
produce a metric as a product of the vierbein gauge fields. To simplify this
model and keep track of the Higgs interactions, we shall consider instead the
coupling of the metric tensor to a gauge theory based on $SP(4)$ spontaneously
broken to $SL(2,C)$ through the Higgs mechanism. In this way the structure of
the massive spin-2 field coupled to a massless graviton will be transparent.

The plan of this paper is as follows. In section two the action representing
the interaction of a metric tensor coupled to $SP(4)$ gauge field and a Higgs
multiplet is constructed. In section three the action is expanded in terms of
$SL(2,C)$ representations of the $SP(4)$ fields. In section four the action is
analyzed in the unitary gauge. In section five the analysis of the action is
done in a non-unitary gauge and the degrees of freedom identified. Section six
is the conclusion.

\section{The action for graviton coupled to a massive spin-2 field}

Consider a metric tensor $g_{\mu\nu}$ on a manifold $M$ and the
Einstein-Hilbert action associated with it%
\[
I_{g}=\frac{1}{4\kappa^{2}}\int\limits_{M}d^{4}x\sqrt{g}R\left(  g\right)  ,
\]
where $M_{Pl}=\frac{1}{\kappa}.$ Consider also the gauge group $SP(4)$ and the
gauge field $W_{\mu}$ associated with it. This can be expanded in terms of the
$SL(2,C)$ subgroup \cite{ch}, \cite{css}%
\[
W_{\mu\alpha}^{\quad\beta}=\left(  L_{\mu}^{a}\left(  \frac{i}{2\kappa_{0}%
}\gamma_{a}\right)  +\frac{1}{4}B_{\mu}^{ab}\gamma_{ab}\right)  _{\alpha
}^{\beta},
\]
which satisfies the symmetry condition
\[
\left(  W_{\mu}C\right)  _{\alpha\beta}=\left(  W_{\mu}C\right)  _{\beta
\alpha},
\]
where $C$ is the charge conjugation matrix. It also satisfies the reality
condition
\[
\gamma_{0}W_{\mu}^{\dagger}\gamma_{0}=-W_{\mu},
\]
implying that $L_{\mu}^{a}$ and $B_{\mu}^{ab}$ are real. The $SP(4)$ gauge
transformation of the gauge field is given by
\[
W_{\mu}\rightarrow\Omega W_{\mu}\Omega^{-1}+\Omega\partial_{\mu}\Omega^{-1},
\]
where the gauge parameter $\Omega$ can be expanded in terms of the $SL(2,C)$
components%
\[
\Omega=\exp\left(  \frac{i}{2}\omega^{a}\gamma_{a}+\frac{1}{4}\omega
^{ab}\gamma_{ab}\right)  .
\]
The component form of the infinitesimal gauge transformations read
\begin{align*}
\delta L_{\mu}^{a} &  =-\kappa_{0}\left(  \partial_{\mu}\omega^{a}+B_{\mu
}^{ab}\omega_{b}\right)  +\omega^{ab}L_{\mu b},\\
\delta B_{\mu}^{ab} &  =-\left(  \partial_{\mu}\omega^{ab}+B_{\mu}^{ac}%
\omega_{c}^{\,\,b}-B_{\mu}^{bc}\omega_{c}^{\,\,a}\right)  -\frac{1}{\kappa
_{0}}\left(  \omega^{a}L_{\mu}^{b}+\omega^{b}L_{\mu}^{a}\right)  .
\end{align*}
The gauge covariant field strengths are defined by%
\[
W_{\mu\nu}=\partial_{\mu}W_{\nu}-\partial_{\nu}W_{\mu}+\left[  W_{\mu},W_{\nu
}\right]  ,
\]
which transforms as $W_{\mu\nu}\rightarrow\Omega W_{\mu\nu}\Omega^{-1}.$ This
can be resolved into%
\[
W_{\mu\nu}=L_{\mu\nu}^{a}\left(  \frac{i}{2\kappa_{0}}\gamma_{a}\right)
+\frac{1}{4}B_{\mu\nu}^{ab}\gamma_{ab},
\]
where
\begin{align*}
L_{\mu\nu}^{a} &  =\partial_{\mu}L_{\nu}^{a}-\partial_{\nu}L_{\mu}^{a}+B_{\mu
}^{ab}L_{\nu b}-B_{\nu}^{ab}L_{\mu b},\\
B_{\mu\nu}^{ab} &  =\partial_{\mu}B_{\nu}^{ab}-\partial_{\mu}B_{\nu}%
^{ab}+B_{\mu}^{ac}B_{\nu c}^{\quad b}-B_{\nu}^{ac}B_{\mu c}^{\quad b}-\frac
{1}{\kappa_{0}^{2}}\left(  L_{\mu}^{a}L_{\nu}^{b}-L_{\nu}^{a}L_{\mu}%
^{b}\right)  .
\end{align*}
Now introduce the Goldstone field $G_{\alpha}^{\beta}$ in the antisymmetric
representation of $SP(4):$
\[
\left(  GC\right)  _{\alpha\beta}=-\left(  GC\right)  _{\beta\alpha},
\]
also subject to the reality and tracelessness conditions
\[
\gamma_{0}G^{\dagger}\gamma_{0}=-G,\qquad G_{\alpha}^{\alpha}=0.
\]
Therefore we can decompose $G$ in the form%
\[
G_{\alpha}^{\beta}=\left(  \varphi\left(  i\gamma_{5}\right)  -v_{a}\left(
\gamma_{a}\gamma_{5}\right)  \right)  _{\alpha}^{\beta},
\]
where $\varphi$ and $v_{a}$ are real fields. In order to isolate the massive
graviton degrees of freedom, we can eliminate one scalar degree of freedom by
imposing a gauge invariant constraint on the multiplet $G$ \cite{ch78}
\[
Tr\left(  G^{2}\right)  =-4a^{2},
\]
which in component form reads
\[
\varphi^{2}+v_{a}v^{a}=a^{2}.
\]
The gauge transformation of $G$ is $G\rightarrow\Omega G\Omega^{-1},$ so that
the covariant derivative is given by%
\[
\nabla_{\mu}G=\partial_{\mu}G+\left[  W_{\mu},G\right]  ,
\]
transforming as $\nabla_{\mu}G\rightarrow\Omega\nabla_{\mu}G\Omega^{-1}.$ The
simplest action for the fields $W_{\mu}$ and $G$ and not involving the
space-time metric is
\begin{align*}
I_{W-G} &  =\int\limits_{M}d^{4}x\epsilon^{\mu\nu\kappa\lambda}Tr\left(
\frac{\alpha}{32a^{3}}\left(  G\nabla_{\mu}G\nabla_{\nu}G+\nabla_{\mu}%
G\nabla_{\nu}GG\right)  W_{\kappa\lambda}\right.  \\
&  \hspace{1in}\left.  +\frac{\beta}{96a^{5}}G\nabla_{\mu}G\nabla_{\nu}%
G\nabla_{\kappa}G\nabla_{\lambda}G\right)  .
\end{align*}
Notice that both terms with the $\alpha$ coefficients are needed for the
action to be Hermitian. One can also add to this action the term%
\[
\int\limits_{M}d^{4}x\epsilon^{\mu\nu\kappa\lambda}Tr\left(  GW_{\mu\nu
}W_{\kappa\lambda}\right)  ,
\]
but in the unitary gauge this will only change, apart from adding a
Gauss-Bonnet topological term, the coefficients of the terms already present.

The next step is to add a mixing terms between the two sectors. To do this
define the field \cite{ags}%
\[
H_{\mu\nu}=g_{\mu\nu}+\frac{\kappa_{0}^{2}}{4a^{2}}Tr\left(  \nabla_{\mu
}G\nabla_{\nu}G\right)  ,
\]
and add the interaction term%
\[
I_{g-H}=m^{4}\int\limits_{M}d^{4}x\sqrt{g}g^{\mu\rho}g^{\nu\sigma}\left(
H_{\mu\nu}H_{\rho\sigma}+(b-1)H_{\mu\rho}H_{\nu\sigma}\right)  .
\]

It is important to note that the multiplet $G$ has the following gauge
transformations in component form%
\begin{align*}
\delta\varphi &  =-\omega^{a}v_{a},\\
\delta v_{a}  &  =\omega_{a}\varphi+\omega_{ab}v^{b}.
\end{align*}
This implies that by an appropriate choice of $\omega_{a}$ it is possible to
use the unitary gauge
\[
v_{a}=0.
\]
so that the constraint simplifies to $\varphi^{2}=a^{2}$, and corresponds to a
non-linear realization of the symmetry breaking from $SP(4)$ to $SL(2,C)$ \cite{zu}

\section{The action in component form}

To derive the action in component form we first write%
\[
\nabla_{\mu}G=\nabla_{\mu}\varphi\left(  i\gamma_{5}\right)  -\nabla_{\mu
}v^{a}\gamma_{a}\gamma_{5},
\]
where
\begin{align*}
\nabla_{\mu}\varphi &  =\partial_{\mu}\varphi-\frac{1}{\kappa_{0}}L_{\mu}%
^{a}v_{a},\\
\nabla_{\mu}v^{a}  &  =\partial_{\mu}v^{a}+B_{\mu}^{ab}v_{b}+\frac{1}%
{\kappa_{0}}L_{\mu}^{a}\varphi.
\end{align*}
The action is then given by%
\begin{align*}
I  &  =\frac{1}{4\kappa^{2}}\int\limits_{M}d^{4}x\sqrt{g}R\left(  g\right) \\
&  +\frac{\beta}{24a^{5}}\int\limits_{M}d^{4}x\epsilon^{\mu\nu\kappa\lambda
}\epsilon_{abcd}\left(  \varphi\nabla_{\mu}v^{a}\nabla_{\nu}v^{b}%
\nabla_{\kappa}v^{c}\nabla_{\lambda}v^{d}-4\nabla_{\mu}\varphi v^{a}%
\nabla_{\nu}v^{b}\nabla_{\kappa}v^{c}\nabla_{\lambda}v^{d}\right) \\
&  -\frac{\alpha}{16a^{3}}\int\limits_{M}d^{4}x\epsilon^{\mu\nu\kappa\lambda
}\epsilon_{abcd}\left(  \left(  \varphi\nabla_{\mu}v^{a}\nabla_{\nu}%
v^{b}-2\nabla_{\mu}\varphi v^{a}\nabla_{\nu}v^{b}\right)  B_{\kappa\lambda
}^{cd}+\frac{2}{\kappa_{0}}v^{a}\nabla_{\mu}v^{b}\nabla_{\nu}v^{c}%
L_{\kappa\lambda}^{d}\right) \\
&  +m^{4}\int\limits_{M}d^{4}x\sqrt{g}g^{\mu\rho}g^{\nu\sigma}\left(
H_{\mu\nu}H_{\rho\sigma}+\left(  b-1\right)  H_{\mu\rho}H_{\nu\sigma}\right)
,
\end{align*}
where
\[
H_{\mu\nu}=g_{\mu\nu}-\frac{\kappa_{0}^{2}}{a^{2}}\left(  \nabla_{\mu}%
\varphi\nabla_{\nu}\varphi+\nabla_{\mu}v^{a}\nabla_{\nu}v_{a}\right)  .
\]

\section{The action in unitary gauge}

It is easy to analyze the action in the unitary gauge $v^{a}=0$ as this
implies
\begin{align*}
\varphi &  =a,\\
\nabla_{\mu}\varphi &  =0,\\
\nabla_{\mu}v^{a} &  =\frac{a}{\kappa_{0}}L_{\mu}^{a},\\
H_{\mu\nu} &  =g_{\mu\nu}-l_{\mu\nu},
\end{align*}
where $l_{\mu\nu}=L_{\mu}^{a}L_{\nu a}.$ In this case the action simplifies to%
\begin{align*}
I &  =\frac{1}{4\kappa^{2}}\int\limits_{M}d^{4}x\sqrt{g}R\left(  g\right)
+\frac{\left(  3\alpha-\beta\right)  }{\kappa_{0}^{4}}\int\limits_{M}%
d^{4}x\det L_{\mu}^{a}\\
&  -\frac{\alpha}{16\kappa_{0}^{2}}\int\limits_{M}d^{4}x\epsilon^{\mu\nu
\kappa\lambda}\epsilon_{abcd}L_{\mu}^{a}L_{\nu}^{b}\,R_{\kappa\lambda}%
^{cd}\left(  B\right)  \\
&  +m^{4}\int\limits_{M}d^{4}x\sqrt{g}g^{\mu\rho}g^{\nu\sigma}\left(
h_{\mu\nu}h_{\rho\sigma}-(1-b)h_{\mu\rho}h_{\nu\sigma}\right)  ,
\end{align*}
where
\begin{align*}
h_{\mu\nu} &  =\left(  g_{\mu\nu}-l_{\mu\nu}\right)  ,\\
R_{\mu\nu}^{ab}\left(  B\right)   &  =\partial_{\mu}B_{\nu}^{ab}-\partial
_{\mu}B_{\nu}^{ab}+B_{\mu}^{ac}B_{\nu c}^{\quad b}-B_{\nu}^{ac}B_{\mu
c}^{\quad b}.
\end{align*}
We can write
\[
\epsilon^{\mu\nu\kappa\lambda}\epsilon_{abcd}\,L_{\mu}^{a}L_{\nu}%
^{b}\,R_{\kappa\lambda}^{cd}\left(  B\right)  =-4\det L\,L_{a}^{\mu}L_{b}%
^{\nu}\,R_{\mu\nu}^{ab}\left(  B\right)  ,
\]
where we have defined $L_{\mu}^{a}L_{a}^{\nu}=\delta_{\mu}^{\nu}.$ It is also
well known that when the field $B_{\mu}^{ab}$ is eliminated by its equations
of motion the above term reduces to%
\[
\frac{\alpha}{4\kappa_{0}^{2}}\int\limits_{M}d^{4}x\det L\,R\left(  l\right)
\]
where $R\left(  l\right)  $ is the curvature of the metric tensor $l_{\mu\nu
}.$ Without any loss in generality we can set $\alpha=1.$ This system is known
to give the coupling of a massive graviton to a massless graviton with the
Fierz-Pauli choice $b=0.$ The massive graviton has mass of the order
$\kappa_{0}m^{2}$, which can be arranged to be small by an appropriate choice
of $m.$ The choice $b=0$ is not stable under quantum corrections and the
propagator does not have a smooth $m\rightarrow0$ limit. When $b\neq0$ there
is a ghost mode in the tensor $l_{\mu\nu}$ which will propagate \cite{bd}. To
find the order of the quantum corrections it is essential to the study the
coupled system in other gauges, and to separate the six degrees of freedom of
the massive tensor into two massless tensor polarizations of $l_{\mu\nu}$, the
two vector polarizations in $v^{a}$ and  two scalar modes.

\section{The non-unitary gauge}

To examine the degrees of freedom in a different gauge, we shall keep the
field $v^{a}$ present, and use the gauge degrees of freedom $\omega^{a}$ to
impose a gauge choice of the form
\[
g^{\mu\nu}\left(  \partial_{\mu}L_{\nu}^{a}+B_{\mu}^{ab}L_{\nu}^{b}\right)
=0,
\]
or something equivalent. This will guarantee that the only propagating degrees
of freedom present in $L_{\mu}^{a}$ are the two tensor degrees of helicities
$+2$ and $-2$. This is easy to see because the system for $L_{\mu}^{a}$ is
identical to the massless graviton with the only difference being that the
diffeomorphism parameters $\zeta^{\mu}$ are used instead of the parameters
$\omega^{a}$ to impose the above gauge condition. From the explicit form of
the action it is evident that all derivatives on the field $v^{a}$ coming from
the terms with coefficients $\alpha$ and $\beta$ are antisymmetrized, implying
that these terms do not give kinetic energy for $v^{a}.$ The only terms that
contain second order derivatives for $v^{a}$ come from the mass mixing term
with coefficient $m^{4}$. \ To see this we first express the field $H_{\mu\nu
}$ in terms of the component fields%
\begin{align*}
H_{\mu\nu} &  =g_{\mu\nu}-\frac{\varphi^{2}}{a^{2}}l_{\mu\nu}-\frac{\kappa
_{0}\varphi}{a^{2}}\left(  \nabla_{\mu}v_{\nu}+\nabla_{\nu}v_{\mu}\right)
-\frac{1}{a^{2}}\left(  v_{\mu}-\kappa_{0}\partial_{\mu}\varphi\right)
\left(  v_{\nu}-\kappa_{0}\partial_{\nu}\varphi\right)  \\
&  -\frac{\kappa_{0}^{2}}{a^{2}}l^{\rho\sigma}\nabla_{\mu}v_{\rho}\nabla_{\nu
}v_{\sigma}+\cdots
\end{align*}
where $v_{\mu}=L_{\mu}^{a}v_{a}$ and $\nabla_{\mu}v_{\nu}=\partial_{\mu}%
v_{\nu}-\Gamma_{\mu\nu}^{\rho}\left(  l\right)  v_{\rho}$ and we have used the
relations
\begin{align*}
-\left(  D_{\mu}L_{\nu}^{a}+D_{\nu}L_{\mu}^{a}\right)   &  =2\Gamma_{\mu\nu
}^{\rho}\left(  l\right)  v_{\rho}+\cdots\\
D_{\mu}v_{a}D_{\nu}v^{a} &  =l^{\rho\sigma}\nabla_{\mu}v_{\rho}\nabla_{\nu
}v_{\sigma}+\cdots
\end{align*}
after substituting the $B_{\mu}^{ab}$ equation of motion. The fields $\varphi$
and $v_{a}$ are related by a constraint, and it is possible to express
$\varphi=\sqrt{a^{2}-v_{a}v^{a}}$ or simply constrain one degree of freedom in
$v_{a}.$ It is then clear that to leading order%
\[
H_{\mu\nu}=h_{\mu\nu}-\frac{\kappa_{0}}{a}\left(  \nabla_{\mu}v_{\nu}%
+\nabla_{\nu}v_{\mu}\right)  +\cdots
\]
Then the mass mixing term gives, to lowest order, the following contribution
to the kinetic energy of $v_{a}$%
\begin{align*}
&  \kappa_{0}^{2}m^{4}\int\limits_{M}d^{4}x\sqrt{g}g^{\mu\rho}g^{\nu\sigma
}\left(  \left(  \partial_{\mu}v_{\nu}+\partial_{\nu}v_{\mu}\right)  \left(
\partial_{\rho}v_{\sigma}+\partial_{\sigma}v_{\rho}\right)  \right.  \\
&  \hspace{1.5in}\left.  -(1-b)\left(  \partial_{\mu}v_{\rho}+\partial_{\rho
}v_{\mu}\right)  \left(  \partial_{\nu}v_{\sigma}+\partial_{\sigma}v_{\rho
}\right)  \right)  .
\end{align*}
After integration by parts and setting $b=0$, this can be rewritten in the
form
\[
\kappa_{0}^{2}m^{4}\int\limits_{M}d^{4}x\sqrt{g}g^{\mu\rho}g^{\nu\sigma
}\left(  \left(  \partial_{\mu}v_{\nu}-\partial_{\nu}v_{\mu}\right)  \left(
\partial_{\rho}v_{\sigma}-\partial_{\sigma}v_{\rho}\right)  \right)  .
\]
This describes the two spin-1 vector polarizations of the massive spin-2
field. There is a  spin-0 polarization which can represented by the mass term
of $v_{a}$ in addition to  the scalar field $\varphi.$ This  is so because
three components of the vector $v_{a}$ propagate, so the constraint can be
thought to be a restriction on the fourth component of $v_{a}.$ In this way
$\varphi$ is represented as
\[
\varphi=a+\overline{\varphi},
\]
where $\overline{\varphi}$ are fluctuations. Although $\varphi$ does not have
a direct kinetic term it occurs as a scaling factor for the curvature scalar
$R(l).$ Isolating the relevant contributions we have%
\[
\int\limits_{M}d^{4}x\det L_{\mu}^{a}\,\varphi^{3}\left(  \frac{1}{4\kappa
_{0}^{2}}R\left(  l\right)  +\frac{\beta-3\alpha}{\alpha\kappa_{0}^{4}%
}\right)  ,
\]
where we have used the constraint on $\varphi$ and $v_{a}$ to simplify
$\beta\varphi^{3}\left(  \varphi^{2}+v_{a}v^{a}\right)  $ to $\beta
a^{2}\varphi^{3}.$ By a Weyl scaling of the field $L_{\mu}^{a}$ the field
$\varphi$ acquires a kinetic energy term, just as the dilaton in string
theory, or when gravity is compactified to lower dimensions. By substituting
the constraint one finds that the mass term for $v_{a}$ is
\[
-\frac{3\left(  \beta-3\alpha\right)  }{2a^{2}\kappa_{0}^{4}}v_{a}v^{a}.
\]
To conclude, in the unitary gauge, the tensor $h_{\mu\nu}$ has six degrees of
freedom, five for the massive graviton coupled to an additional scalar degree
of freedom. In the non-unitary gauge, there are two tensor polarizations of
helicities $\pm2$ (corresponding to the tensor $l_{\mu\nu}$), two
polarizations of helicities $\pm1$ corresponding to the transverse components
of $v_{a}$ and two scalar polarization of spin-0 corresponding to the field
$\varphi$ and the longitudinal component of $v_{a}.$ The ill behavior of the
massive graviton propagator can be avoided by considering instead of the
$g_{\mu\nu},$ $l_{\mu\nu}$ system, the coupled constrained system of
$\ $\ $g_{\mu\nu},$ $L_{\mu}^{a},$ $\varphi$ and $v_{a}$. The instability of
the Fierz-Pauli choice of $b=0$ would occur at the quantum level, but as
explained in \cite{ags}, the corrections would occur at a cut-off energy,
where the ghost mode would start to propagate. At energies much lower than the
cut-off scale, the corrections could be ignored, and the system is well behaved.

\section{Conclusions}

We have shown that it is possible to formulate an action for a massless
graviton interacting with a massive spin-2 field as a theory obtained by
coupling a metric tensor to a gauge theory of $SP(4)$ spontaneously broken to
$SL(2,C).$ The symmetry breaking is done through a Higgs multiplet containing
a scalar field and a vector field related by a constraint, employing a
non-linear realization for the symmetry breaking. The action consists of three
parts. The first part is the Einstein-Hilbert action for the metric tensor
$g_{\mu\nu}$ on a space-time manifold $M.$ The second part is metric
independent and $SP(4)$ gauge invariant. The third is a mixing term between
the metric and the gauge sectors. In the unitary gauge the Goldstone vector
field $v_{a}$ is set to zero and the field $\varphi$ to a constant. The action
reduces to the Fierz-Pauli form of a massless graviton interacting with a
massive tensor. In the non-unitary gauge, the degrees of freedom of the
massive spin-2 field and the scalar are given by the tensor polarizations of
helicities $\pm2$ in $L_{\mu}^{a}$, the vector polarizations of helicities
$\pm1$ in $v_{a}$ and two scalars \ of helicities $0$ in $\varphi$ and $v_{a}%
$. In this form the propagators of the separate modes are well behaved and
have a smooth $m\rightarrow0$ limit. Quantum corrections to the Fierz-Pauli
choice of the mass term would be damped by the cut-off energy. At energies
much below the cut-off the action is well behaved. What remains to be seen is
an explicit computation to show how the massless limit is attained. More
importantly is to have a variant of this action where the system is treated
more symmetrically with the two tensors corresponding to the two metrics on
the separate sheets of two membranes. This can be done either by considering a
purely metric theory as in \cite{ags}, or by considering a gauge theory based
on the gauge group $SP(4)\times SP(4)$ broken to $SL(2,C).$ This is indeed
possible as shown in \cite{css}, but is complicated by the fact that the Higgs
field transforms under $SP(4)\times SP(4)$ and has $16$ components requiring
the introduction of a good number of constraints. The resulting model is very
similar to what is described here and is fairly straightforward in the unitary
gauge. But the analysis is more complicated in the non-unitary gauge as one
has to keep track of all independent components of the Goldstone fields.

\section{Acknowledgments}

I would like to thank the Alexandar von Humboldt Foundation for support
through a research award. I would also like to thank Slava Mukhanov for
hospitality at the Ludwig-Maxmilians University in M\"{u}nich where part of
this work was done.

\end{document}